\documentclass[vecphys]{svmult}

\usepackage{makeidx}         % allows index generation
\usepackage{graphicx}        % standard LaTeX graphics tool
\usepackage{multicol}        % used for the two-column index
\usepackage[bottom]{footmisc}% places footnotes at page bottom
\makeindex             % used for the subject index
\begin{document}

\title*{Resolving the inner active accretion disk around the Herbig Be star
  MWC\,147 with VLTI/MIDI+AMBER spectro-interferometry}
\titlerunning{Resolving the Herbig Be star MWC\,147 with VLTI spectro-interferometry}
\author{S.~Kraus\and
Th.~Preibisch\and
K.~Ohnaka}
\institute{Max-Planck-Institut f\"ur Radioastronomie, Auf dem H\"ugel 69,
  53121 Bonn, Germany
\texttt{skraus@mpifr-bonn.mpg.de}}

\maketitle

\begin{abstract}
We studied the geometry of the inner (AU-scale) circumstellar environment around
the Herbig~Be star MWC\,147.
Combining, for the first time, near- (NIR, $K$~band) and mid-infrared (MIR,
$N$~band) spectro-interferometry on a Herbig star, our VLTI/MIDI and AMBER 
data constrain not only the geometry of the brightness distribution, but also
the radial temperature distribution in the disk. 
For our detailed modeling of the interferometric data and the spectral energy
distribution (SED), we employ 2-D radiation transfer
simulations, showing that passive irradiated Keplerian dust 
disks can easily fit the SED, but predict much lower visibilities than
observed.
Models of a Keplerian disk with emission from an optically thick inner
gaseous accretion disk (inside the dust sublimation zone), however, yield a
good fit of the SED and simultaneously reproduce the observed
NIR and MIR visibilities. We conclude that the NIR continuum emission from
MWC\,147 is dominated by accretion luminosity emerging from an optically thick
inner gaseous disk, while the MIR emission also contains strong contributions
from the outer dust disk. 
\end{abstract}

\section{Introduction and observations}
\label{sec:1}

Herbig Ae/Be (HAeBe) stars are intermediate-mass objects which are still
accreting material, probably via a circumstellar disk composed of gas and dust.
Understanding the structure of these disks and the processes through which
they interact with the central star is critical for our
understanding of the formation process of stars.
Since, until recently, the spatial scales of the inner circumstellar
environment (a few AU) were not accessible to infrared imaging observations,
conclusions drawn on the 3-D geometry of the circumstellar material were
mostly based on the modeling of the SED (e.g.\
Hillenbrand et al.\ \cite{hil92}).
However, these fits are known to be ambiguous (e.g.\ Men'shchikov \&
Henning \cite{men97}) and have to be complemented with spatial information,
as provided by infrared interferometry. 

The first systematic studies of HAeBes using the technique of infrared
long-baseline interferometry  
have revealed that for most HAeBes the NIR
size correlates with the stellar luminosity $L$ following a simple $R
\propto L^{1/2}$ law.  This suggests that the NIR continuum  emission mainly
traces hot dust at the inner sublimation radius. 
However, for more luminous Herbig Be stars, the NIR-emitting structure is
more compact than predicted by the size-luminosity relation (Monnier et al.\ \cite{mon02}).

To further investigate the origin of the ``undersized'' Herbig~Be star
disks, we observed the Herbig~Be star MWC\,147 with the VLTI. First
infrared interferometric observations of this star by Akeson et al.\
\cite{ake00} indicated that the NIR-emitting region is surprisingly
compact ($\sim0.7$~AU, assuming a uniform disk profile).
In the course of three ESO open time programmes and using the 8.2~m UT
telescopes, we obtained seven VLTI/MIDI measurements and one VLTI/AMBER
measurement.
The MIDI observations cover baseline lengths between 39 and 102~m.  
From the AMBER data, one wavelength-dependent visibility, corresponding to a
101~m baseline, could be extracted.
For our modeling of MWC\,147, we adopt the stellar parameters by
{Hern{\'a}ndez} et al.\ \cite{her04}, namely a spectral type of B6, a
distance of 800~pc, a bolometric luminosity of 1550~$L_{\odot}$, a mass of
6.6~$M_{\odot}$, and a stellar radius of 6.63~$R_{\odot}$.  

\section{The power of joint NIR/MIR spectro-interferometry}
\label{sec:2}

\begin{figure}[b]
\centering
\includegraphics[height=2.9cm]{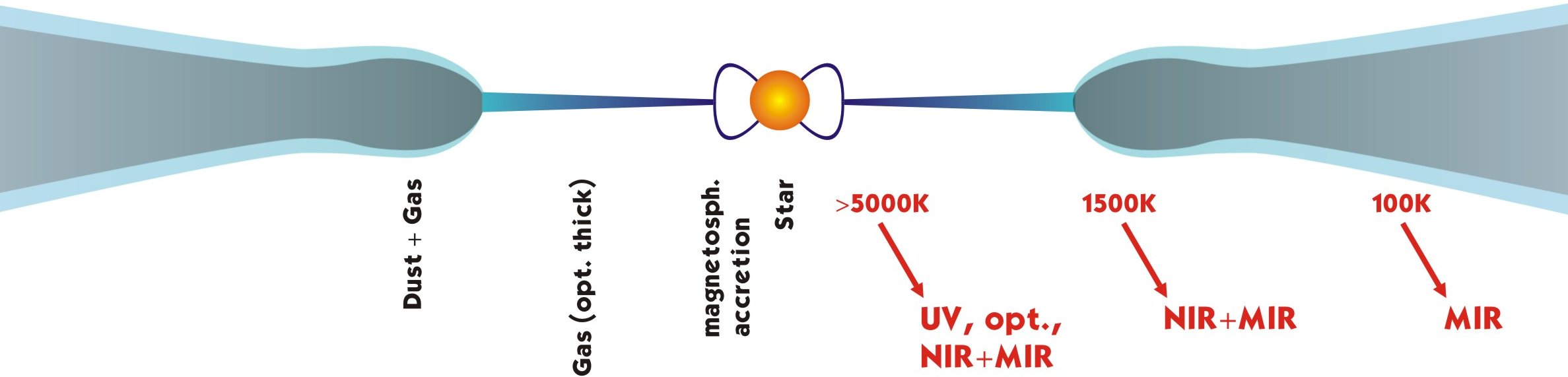}
\caption{Illustration of the inner environment of HAeBe stars.  Due to the
  different temperatures, spectro-interferometry can disentangle
  multiple emission components. }
\label{fig:1}
\end{figure}

The VLTI instruments AMBER and MIDI combine the high spatial resolution
achievable with infrared interferometry with spectroscopic capabilities,
measuring the fringe visibility as a function of wavelength.
As circumstellar disks exhibit a temperature gradient, different spectral
channels trace different spatial regions.  Therefore, spectro-interferometric
observations, which cover a sufficiently large wavelength range, can constrain
not only the disk geometry, but also the radial temperature profile of the
disk.
For the investigation on YSO disks, the NIR and MIR wavelength regimes are
particularly well suited since the NIR wavelength regime ($\sim
2~\mu$m) is most sensitive to the thermal emission from dust located at the
dust sublimation radius ($T \approx  1500$~K, a few AU from the star), while
MIR wavelengths ($\sim 10~\mu$m) trace dust with a temperature of several
hundred Kelvin, located a few $10$~AU from the star (see Fig.~\ref{fig:1}). 

\section{Radiative transfer modeling}
\label{sec:3}

\begin{figure}[h]
\vspace{-3mm}
\begin{flushleft}
\resizebox{6.7cm}{!}{\includegraphics[angle=0]{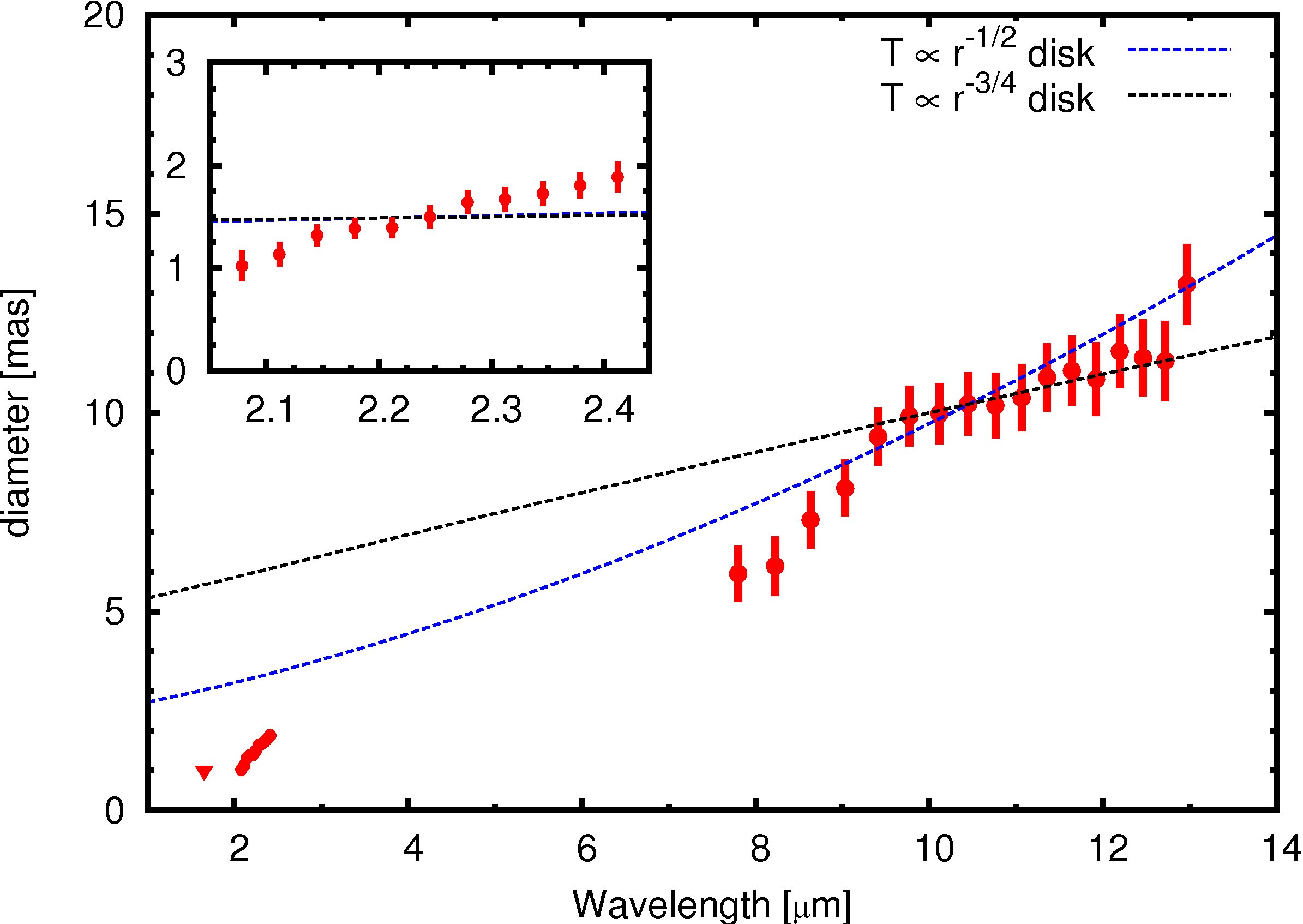}}
\end{flushleft}
\vspace*{-5.5cm}
\begin{flushright}
\begin{minipage}{4.9cm}
\caption{
  Comparing the wavelength-dependent disk size of MWC\,147 (as derived from
  one of our VLTI/AMBER+MIDI measurements) with the wavelength-dependent disk size
  predicted by standard temperature power-law disk models, we find that these
  models cannot reproduce our measurements.  In this figure, the disk model
  was scaled to match the measured MIR size.
}
\label{fig:2}
\end{minipage}
\end{flushright}
\vspace{-7.1mm}
\end{figure}

As a first step of analysis, we compared the interferometric data to
commonly used analytic disk models with a simple temperature power-law
($T(r) \propto r^{-\alpha}$, with $\alpha=3/4$ or $1/2$). 
Using the assumption that each disk annulus radiates as a blackbody, % and a
we can compute the wavelength-dependence of the disk size corresponding to
these analytic models and find that they
cannot reproduce the measured NIR and MIR-sizes simultaneously (see Fig.~\ref{fig:2}).

Therefore, we applied a more sophisticated modeling approach using the
\textit{mcsim\_mpi} 2-D radiative transfer code (Ohnaka et~al.\
\cite{ohn06}). 
For each model, we first check the agreement with the 
SED of MWC\,147 and then fit the spectro-interferometric
visibilities (see Kraus et al.\ \cite{kra07} for details). 
The dust density distribution of the accretion disk in our models resembles a
flared, Keplerian-rotating disk with a radial density distribution of 
$\rho(r)\propto r^{-3/2}$, which extends from the dust sublimation radius
to 100~AU.
In order to reproduce the shape of the SED, we found that,
in addition to the disk, an extended envelope is required,
for which we use the radial density distribution 
$\rho(r)\propto r^{-1/2}$.

\begin{figure}[t]
\centering
\includegraphics[width=12.1cm]{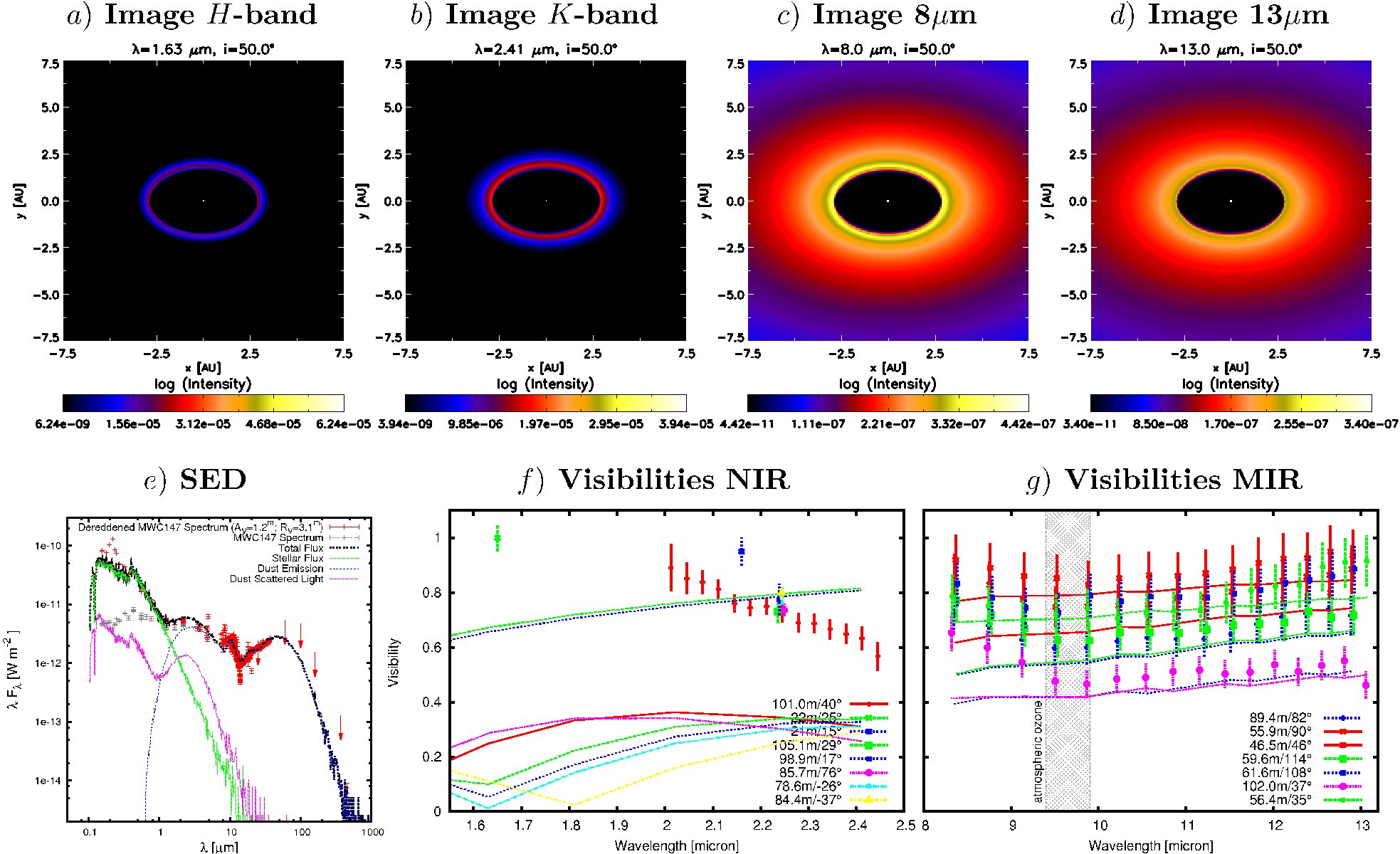}
\caption{Model images {\it (a-d)}, SED {\it (e)}, and
  NIR/MIR visibilities {\it {f-g}} corresponding to our best-fit radiative
  transfer image of an irradiated dust disk geometry.  The poor agreement
  between measured and model visibilities ($\chi^2_r \approx 26$) indicates
  that passive disk models can be ruled out. }
\label{fig:3}
\end{figure}

Fig.~\ref{fig:3} shows model images, the SED, as well as the visibilities
corresponding to our best-fit model of a passive, irradiated accretion disk.
Although irradiated disks are able to reproduce the SED of MWC\,147, 
they predict visibilities which are much smaller than the measured
visibilities ($\chi^2_r \approx 26$) and are therefore in strong conflict with
our interferometric measurements.

In passive circumstellar disks, the infrared emission is generally assumed to
originate almost entirely from dust; the emissivity of the inner, dust-free
gaseous part of the disk, at radii smaller than the dust sublimation radius,
is negligible. In an actively accreting disk, on the other hand, viscous
dissipation of energy in the inner dust-free gaseous part of the accretion
disk can heat the gas to high temperatures and give rise to significant
amounts of infrared emission from optically thick gas. The inner edge of this
gas accretion disk is expected to be located a few stellar radii above the
stellar surface, where the hot gas is thought to be channeled towards the star
via magnetospheric accretion columns.
While the magnetospheric accretion columns are too small to be resolved in our
interferometric data (3~$R_{\star}$ correspond to 0.09~AU or 0.12~mas), 
infrared emission from hot gas between the dust sublimation radius 
and the stellar surface should be clearly distinguishable from the thermal
emission of the dusty disk  due to the different temperatures of these
components and the resulting characteristic slope in the NIR- and
MIR-visibilities.  

\begin{figure}[t]
\centering
\includegraphics[width=12.1cm]{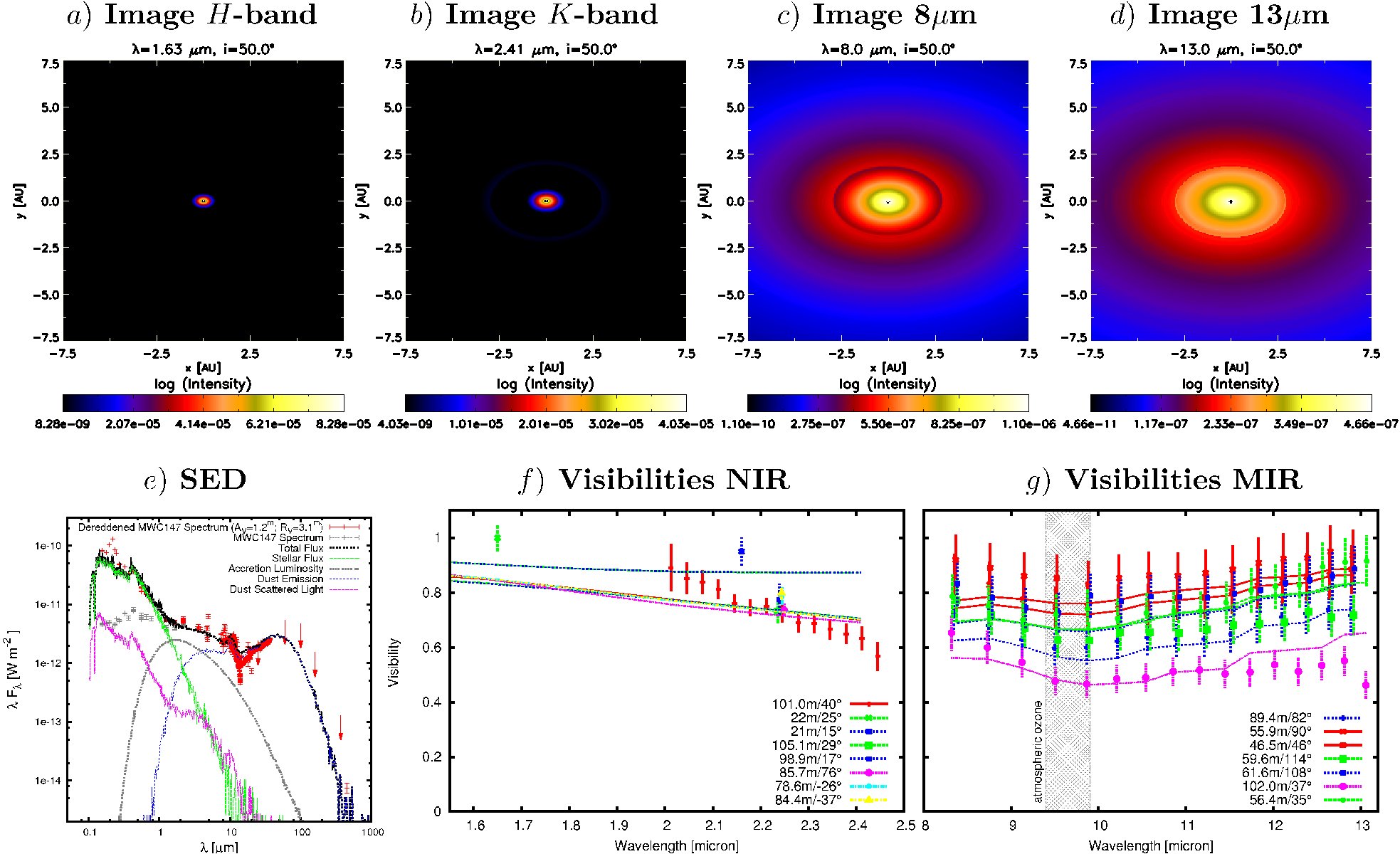}
\caption{Observables corresponding to our best-fit radiative transfer model with
  optically-thick inner gas disk (similar to Fig.~\ref{fig:3}), yielding good
  agreement ($\chi^2_r \approx 1.0$). }
\label{fig:4}
\end{figure}

Since MWC\,147 is a quite strong accretor ($\dot{M}_{\rm acc} \approx
10^{-5}~M_{\odot}$\,yr$^{-1}$; Hillenbrand et~al.\ \cite{hil92}), 
significant infrared emission from the inner gaseous accretion disk is expected.
Muzerolle et al.\ \cite{muz04} found that even for smaller accretion rates,
the gaseous inner accretion disk is several times thinner than the puffed-up 
inner dust disk wall and is optically thick (both in radial as well 
as in the vertical direction).
In order to add the thermal emission from the inner gaseous disk to our
radiative transfer models, we assume the radial temperature power-law by
Pringle \cite{pri81}.  Including the accretion luminosity from an inner
gaseous disk in the model strongly improves the agreement between model
predictions and observed visibilities. 
With a flared disk geometry and an accretion rate of 
$\dot{M}_{\rm acc} = 9 \times 10^{-6}~M_{\odot}$\,yr$^{-1}$, both the SED and the
interferometric visibilities are reproduced reasonably well ($\chi^2_r=1.0$,
see Fig.~\ref{fig:4}).

\section{Conclusions and outlook}
\label{sec:4}

Our VLTI interferometric observations of MWC\,147 constrain, for the first
time, the inner circumstellar environment around a Herbig~Be star over the
wavelength range from 2 to $13~\mu$m.
We find evidence that the NIR emission of MWC\,147 is
dominated by the emission from optically-thick gas located inside the dust
sublimation radius, while the MIR also contains contributions from
the outer, irradiated dust disk.

Our study demonstrates the power of infrared spectro-interferometry
to probe the inner structure of the disks around young stars and to
disentangle multiple emission components.
Future investigations on YSO accretion disks will benefit substantially from
the proposed 2nd~generation VLTI instruments, such as MATISSE, increasing
not only the number of recorded baselines, but also expanding the spectral
coverage to the $L$ and $M$ bands.

\printindex

\begin{thebibliography}{99.}
% Use the following syntax and markup for your references
%

\bibitem{hil92}
{Hillenbrand}, L.~A., {Strom}, S.~E., {Vrba}, F.~J.,  \& {Keene}, J. 1992,
  ApJ, 397, 613

\bibitem{men97} Men'shchikov, A.~B., \& Henning, T.\ 1997, A\&A, 318, 879 

\bibitem{mon02}
{Monnier}, J.~D.,  \& {Millan-Gabet}, R. 2002, ApJ, 579, 694

\bibitem{ake00}
{Akeson}, R.~L., {Ciardi}, D.~R., {van Belle}, G.~T., {Creech-Eakman}, M.~J.,
  \& {Lada}, E.~A. 2000, ApJ, 543, 313

\bibitem{her04}
{Hern{\'a}ndez}, J., {Calvet}, N., {Brice{\~n}o}, C., {Hartmann}, L.,  \&
  {Berlind}, P. 2004, AJ, 127, 1682

\bibitem{ohn06}
{Ohnaka}, K., et~al. 2006, A\&A, 445, 1015

\bibitem{kra07} Kraus, S., Preibisch, T., \& Ohnaka, K.\ 2007, arXiv:0711.4988 

\bibitem{muz04}
{Muzerolle}, J., {D'Alessio}, P., {Calvet}, N.,  \& {Hartmann}, L. 2004, ApJ,
  617, 406

\bibitem{pri81}
{Pringle}, J.~E. 1981, ARA\&A, 19, 137

\end{thebibliography}
\end{document}